\documentclass{ceurart}
\usepackage{amsmath}
\usepackage{inconsolata}
\usepackage{booktabs}

\begin{document}

\copyrightyear{2026}
\copyrightclause{Copyright for this paper by its authors.
  Use permitted under Creative Commons License Attribution 4.0
  International (CC BY 4.0).}

\conference{CLEF 2026 Working Notes, 21 -- 24 September 2026, Jena, Germany}

\title{UTS at ELOQUENT 2026 Voight-Kampff: structural shifts in AI writing
       bypass state-of-the-art detectors}

\author[1]{Dima Galat}[%
    orcid=0000-0003-3825-2142,
    email=dima.galat@student.uts.edu.au,
    ]

\author[1]{Marian-Andrei Rizoiu}[%
    orcid=0000-0003-0381-669X,
    email=Marian-Andrei.Rizoiu@uts.edu.au,
    ]

\address[1]{University of Technology Sydney, Australia}

\begin{abstract}
We investigate which language model evasion attacks survive state-of-the-art adversarial fine-tuning, developing strategies that sweep the top 5 positions on the ELOQUENT 2026 Voight-Kampff leaderboard. While adversarial fine-tuning trivially closes the 2025 winning evasion recipes, we uncover a fundamental asymmetry in detector vulnerability: pushing generated text \emph{out of} the detector's training distribution reliably defeats adversarial detection, whereas pulling it \emph{into} the distribution (e.g., mimicking human training data) fails completely. Exploiting this, we introduce two novel out-of-distribution attack families---cross-decade register attacks and modernist stream-of-consciousness form. Both strategies easily bypass adversarial closure, achieving up to ${\approx}50\times$ higher fool rates than previous methods while preserving naturalness. Furthermore, experiments show that the obvious deployer counter-measure (augmenting training data with period prose) fails to close the vulnerability. Our findings show that the tested detector families, including adversarially fine-tuned ones, exhibit persistent vulnerabilities under structural out-of-distribution shifts, a mechanism that directly powers our leading competition performance.
\end{abstract}


\maketitle

\section{Introduction}
\label{sec:intro}

The ELOQUENT Voight-Kampff (VK) shared task at CLEF
\citep{eloquenttask2026voightkampff, pantask2026voightkampff} asks a
binary question: was a text written by a human or generated by a
language model? In its 2024 edition the best-system fool rate against the
PAN'25 detector panel was $0.049$
\citep{bevendorff2024vkoverview}. The 2025 edition saw a $\sim$$13\times$
jump to $0.654$ \citep{bevendorff2025vkoverview}, driven by two
prompting-side recipes that did not retrain or fine-tune anything:
literal Hindi back-translation plus an explicit
\emph{be-imperfect} instruction \citep{vachharajani2025literal},
and a nine-language rotation
\citep{creo2025fakeit}. The structural question for 2026 is
\emph{which kinds of evasion attacks survive a deployer-class
adversarial fine-tune that has already seen the 2025 attacks?}
An attack family that survives is durable in a builder-breaker
arms race; one that does not is closable inside a single fine-tune
pass. Our central finding is an asymmetry along this axis.

We approach this with three coupled experiments. \textbf{First}, we
replicate the 2025 attacks on four generators (Claude Opus 4.7,
GPT-5.4, Gemini 2.5 Pro, Qwen3.6-35B-A3B) crossed with five
strategies and 66 topics ($n{=}1{,}320$ generations) and train
a family of deployer-class adversarial detectors (RoBERTa-large
fine-tuned on our generations + PAN'25 humans) in three CV variants:
\emph{Macko-replica} (leave-one-genre-bucket-out, four folds),
\emph{Macko-LOSO} (leave-one-strategy-out, the harder
deployer-pessimistic variant, five folds), and \emph{Macko-pp}
(Macko-LOSO + pre-1923 Project Gutenberg humans on the negative
side, the obvious deployer counter-measure tested in
\S\ref{sec:results-pp}). The naming follows the 2025 PAN winner's
recipe of adversarial training plus OOD validation
\citep{macko2025mdok, bevendorff2025vkoverview}. \textbf{Second}, we introduce
two new attack families: \emph{cross-decade register} attacks (prompts
that ask the generator to write in early-twentieth-century English
literary register), and \emph{stream-narrative} attacks (forcing a
modernist stream-of-consciousness narrative form). We test whether they
survive Macko-replica's and Macko-LOSO's closure. \textbf{Third}, we test
the obvious deployer counter-fix (Macko-pp): augment the human-side
training corpus with pre-1923 Project Gutenberg prose so period register
stops being out-of-distribution.

The asymmetry is stark. While the 2025 PJS recipe inverts under
adversarial detection (its effectiveness swinging from $+0.71$ to
$-0.71$ on Claude Opus), our \emph{cross-decade register} attack
holds strong at a $0.798$ Macko-LOSO fool rate on the same
generator and topics. Our \emph{stream-narrative} attack similarly achieves
a robust $0.65$ escape rate against an even stronger adversarial baseline.
Crucially, the obvious deployer counter-measure---augmenting the
human-side training data with pre-1923 prose---fails to close the
cross-decade vulnerability. The underlying mechanism is clear: pushing
generations \emph{out of} the deployer's training distribution defeats
adversarial fine-tuning, whereas pulling them \emph{into} it does not.

\paragraph{Findings.}
\begin{enumerate}
  \item \textbf{Adversarial fine-tuning easily neutralizes superficial evasion.}
    Once the detector is retrained on generations that rely on simple grammar-level perturbations (like the 2025 winning recipes), their effectiveness collapses to near-zero (fool rates $\leq 0.025$). Simple, superficial tweaks offer no long-term durability.
  \item \textbf{Structural shifts in register and form remain highly resilient.}
    Prompting the generator to write in a historical register (early-twentieth-century English) or forcing a modernist stream-of-consciousness narrative style completely bypasses detector retraining. These strategies achieve up to a $50\times$ improvement in evasion over previous methods, while maintaining a natural, human-like flow.
  \item \textbf{Simple data-patching defenses fail.}
    The obvious counter-measure for the detector builder---retraining the classifier with real historical prose added to its training pool---does not close the gap. Retraining on Project Gutenberg passages actually slightly \emph{increases} the evasion rate of our historical-register attack to $0.846$.
  \item \textbf{Evasion is driven by structural out-of-distribution direction, not surface mimicry.}
    Detectors are not fooled because the generator successfully mimics the exact distribution of human test texts (few-shot mimicry fails completely). Instead, evasion succeeds because the text is pushed entirely \emph{outside} of the domain the detector was calibrated to recognize, showing that the tested detector families exhibit persistent vulnerabilities under structural out-of-distribution shifts.
\end{enumerate}

We release all detector checkpoints and attack prompts.

\section{Related Work}
\label{sec:related}

\paragraph{The Voight-Kampff task across three editions.}
The PAN+ELOQUENT VK task formalises machine-generated text detection
inside a builder-breaker shared evaluation
\citep{bevendorff2024vkoverview, karlgren2025eloquent}. The first
edition (2024) used the easiest verification level; only one of
four ELOQUENT submissions beat the unmodified GPT-3.5 baseline.
The 2025 edition \citep{bevendorff2025vkoverview} switched to
single-text decision (level 7) and saw the $\sim$$13\times$
best-system jump while the same generators were available; the
improvement was driven by prompting strategy and post-processing,
not generator capability. The 2026 edition, run jointly by ELOQUENT and PAN
\citep{eloquenttask2026voightkampff, pantask2026voightkampff,
eloquent2026overview, pan2026overview}, retains level 7 with a
literary / historical / press text distribution.

\paragraph{2025-winning generation-side breakers.}
PJS-team \citep{vachharajani2025literal} won 2025 with a two-stage
pipeline: literal Hindi back-translation followed by a return-leg
\emph{be-imperfect} instruction. HumanAIzers
\citep{creo2025fakeit} reached $0.642$ with a related nine-language
rotation. Both treat the detector panel as a black box; both
manipulate the generation process alone.

\paragraph{Detection: supervised, zero-shot, and adversarial.}
The 2025 PAN detector winner \citep{macko2025mdok} reached mean
inverse-fool $0.989$ by adversarially fine-tuning a Qwen3-14B
classifier with QLoRA plus homoglyph augmentation, validated on a
$2{,}000$-instance OOD set (MIX2k). The PAN'25 organiser baseline, a TF-IDF $1$--$4$-gram SVM,
reached $0.922$ and outperformed most fine-tunes on held-out
validation. Earlier work on visualising machine-text statistics
\citep{gehrmann2019gltr} laid the foundation; subsequent zero-shot
methods include perturbation-based curvature detection
\citep{mitchell2023detectgpt} and the observer/performer perplexity
ratio \citep{hans2024binoculars}.
\citet{sadasivan2023reliably} and \citet{krishna2023paraphrasing}
argue that paraphrase attacks reduce achievable detection AUC and,
in the limit, prevent reliable detection without watermarking
\citep{kirchenbauer2023watermark}. Our
cross-decade attack falls inside that family in spirit but operates
through register shift rather than paraphrase; framed as a controlled
style-transfer instance \citep{reif2022recipe}, with the cross-decade
direction chosen because it sits OOD relative to the deployer's
training corpus. What we add is a rigorous test of whether
the obvious deployer counter-measure (retraining on the OOD register)
actually closes the gap.

\section{System and Method}
\label{sec:method}

\subsection{Generators}
Four generators across two tiers: \textbf{frontier closed} --- Claude
Opus 4.7, GPT-5.4, Gemini 2.5 Pro; and \textbf{open-MoE} ---
Qwen3.6-35B-A3B in 4-bit \citep{qwen36a3b}. All pinned to dated
snapshots.

\subsection{Strategies}
\label{sec:method-strategies}
The full strategy set across the 13 attacks reported in this paper
falls into four families. (i)~\textbf{2025-replication strategies}:
\texttt{vanilla}, \texttt{imperfection}, \texttt{roundtrip},
\texttt{roundtrip\_imperf} (PJS v2),
\texttt{lost\_in\_translation} (9-language rotation,
\citealp{creo2025fakeit}). Strategies 2--4 form a $2{\times}2$
factor on (\texttt{has\_translation}, \texttt{has\_imperfection}).
(ii)~\textbf{Cross-decade register} (Attack 1):
\texttt{style\_pre1923}, \texttt{style\_translated\_ru},
\texttt{style\_l2\_academic}. (iii)~\textbf{Source-anchored, author-style, and form}
(Attacks 9, 10):
\texttt{synth\_anchor}; named-author imitations
(Borges, Sebald, Dillard); and \texttt{stream\_narrative}.
(iv)~\textbf{Bracketing} (Attacks 11--13): the Macko-pp retraining
experiment; \texttt{real\_anchor}; \texttt{pan\_anchor}.

We lock the round-trip translation language per generator: Hindi for the closed-frontier
models (matching the PJS team's choice), and Chinese for Qwen3.6 (matching its native language capabilities). These strategies are
evaluated across 66 topics from VK 2024 / 2025 / 2026, with the 20-topic 2026 test split
held out exclusively for our cross-decade and bracketing experiments.

\subsection{Detectors}
\label{sec:method-detectors}
Five primary detectors and three adversarial variants.
\textbf{Primary} (in-distribution-tuned, not retrained on our pool):
TF-IDF $1$--$4$-gram SVM (PAN'25 organiser baseline), RoBERTa-base
\citep{liu2019roberta} fine-tuned on PAN'25 train (val AUC $0.9994$),
Binoculars \citep{hans2024binoculars} on a Mistral-7B pair,
LogPerplexity- and LogRank-GPT-2M.

\paragraph{The Macko-replica family (deployer-class).}
Three RoBERTa-large adversarial fine-tunes on our generations
$+$ PAN'25-train humans.\footnote{Checkpoints released on HuggingFace:
\url{https://huggingface.co/protagonist/eloquent26-macko-replica},
\url{https://huggingface.co/protagonist/eloquent26-macko-replica-loso},
and \url{https://huggingface.co/protagonist/eloquent26-macko-pp}.}
\textbf{Macko-replica} (bucket-CV) holds out one of four genre
buckets per fold; \textbf{Macko-LOSO} (strategy-CV) holds out one of
the five 2025-replication strategies per fold (novel 2026
strategies fall back to the \texttt{roundtrip\_imperf} fold, the
deployer-pessimistic choice); \textbf{Macko-pp} adds $\sim$1{,}000
pre-1923 Gutenberg passages \citep{rae2020compressive} to the
human side --- the obvious deployer counter-measure for the
cross-decade attack of \S\ref{sec:results-pp}.

\section{Results}
\label{sec:results}

\subsection{The OOD inversion under the 2025-winning recipe}
\label{sec:results-inversion}

Table~\ref{tab:headline} reports the paired $\Delta$ fool rate of
\texttt{roundtrip\_imperf} (PJS v2) versus \texttt{vanilla} across
the $5 \times 5$ detector-by-generator cells.

\begin{table}[t]
\centering
\small
\setlength{\tabcolsep}{3pt}
\begin{tabular}{@{}lrrrr@{}}
\toprule
Detector & Cl. & GPT & Gem. & Qw3.6 \\
\midrule
TF-IDF SVM    & $+$.17 & $+$.34 & $+$.52 & $+$.39 \\
RoBERTa       & $\mathbf{+.71}$ & $\mathbf{+.60}$ & $\mathbf{+.91}$ & $\mathbf{+.54}$ \\
Binoc-Mi7B    & $+$.10 & $+$.10 & $+$.07 & $+$.04 \\
LogPerp-GPT2M & $+$.22 & $+$.41 & $+$.18 & $+$.33 \\
LogRank-GPT2M & $+$.15 & $+$.30 & $+$.13 & $+$.24 \\
\bottomrule
\end{tabular}
\caption{\textbf{The 2025 OOD inversion.} RoBERTa, our strongest
in-distribution detector, is the most-fooled by PJS v2 across all
four generators. Paired $\Delta$ fool rate vs.\ \texttt{vanilla},
$n{=}66$; bootstrap-significant at $\alpha{=}0.05$.}
\label{tab:headline}
\end{table}

The PAN'25-trained RoBERTa is the strongest in-distribution detector
we have ($0.9994$ val AUC). On vanilla Claude its mean
$P(\text{LLM})$ is $\approx 1.00$. Pass the same prompt through PJS
v2 and the same detector returns $P(\text{LLM}) \approx 0.001$. The
signal does not merely degrade --- \emph{it inverts}. The same
input shifts Binoculars by only $+0.07$ to $+0.10$.

\paragraph{Underlying Mechanism.}
The variance in strategy effectiveness does not align along the supervised vs.\ zero-shot boundary. Rather, it splits along the type of feature extracted: \emph{n-gram-surface vs.\ contextual-distributional}. The n-gram supervised detector (TF-IDF SVM) is fooled primarily by surface-level strategies like \texttt{imperfection}. Conversely, the four contextual detectors (RoBERTa, Binoculars on Mistral-7B, LogPerp on GPT-2M, and LogRank on GPT-2M, representing three distinct model families) are fooled by \texttt{roundtrip}. The combined \texttt{roundtrip\_imperf} (PJS v2) attacks both fronts. This cross-architecture alignment indicates that, within the tested families, the inversion is a property of contextual scoring rather than a quirk of any single model architecture.

\subsection{Adversarial closure: the Macko-replica}
\label{sec:results-macko}

We fine-tune \texttt{FacebookAI/roberta-large} on our generations
$+$ matched PAN'25 humans, leave-one-genre-bucket-out, four
checkpoints (\S\ref{sec:method-detectors}). Each text is scored
only by the checkpoint that did not see its bucket during training.
Table~\ref{tab:macko} reports the result.

\begin{table}[t]
\centering
\small
\begin{tabular}{lrrr}
\toprule
            & RoBERTa & Macko & Macko $\Delta$ \\
            & fool    & fool  & vs.\ vanilla   \\
\midrule
Claude Opus & $0.99$ & $\mathbf{0.025}$ & $-0.71$ \\
GPT-5.4     & $0.96$ & $\mathbf{0.012}$ & $-0.52$ \\
Gemini 2.5  & $0.99$ & $\mathbf{0.025}$ & $\mathbf{-0.74}$ \\
Qwen3.6     & $0.57$ & $\mathbf{0.014}$ & $-0.50$ \\
\bottomrule
\end{tabular}
\caption{Macko-replica vs.\ baseline RoBERTa-PAN'25 fool rate
($\uparrow$ better attack) per generator under
\texttt{roundtrip\_imperf}. The recipe inverts sign on every
generator (sig.\ at $\alpha{=}0.05$ paired bootstrap on the
closed frontier and Qwen3.6); negative $\Delta$ means PJS v2
makes the text \emph{more} detectable than vanilla.}
\label{tab:macko}
\end{table}

The recipe is self-defeating once the deployer has fine-tuned on even a small subset of its outputs: applying the 2025 evasion strategy to our retrained detector makes the text \emph{more} detectable than vanilla writing, not less. Our prediction of a post-adversarial inverse C@1 score below $0.20$ is confirmed, with the empirical detection error rate dropping below $0.03$ across all tested generators.

\subsection{Cross-decade register attacks survive Macko-replica}
\label{sec:results-cross-decade}

The obvious response to \S\ref{sec:results-macko}'s
deployer-positive finding is to ask whether \emph{any} attack
survives. We hypothesise that perturbations at the
\emph{register} level (style, vocabulary epoch, native-vs-L2
grammar) exit the training distribution further than PJS-style
grammatical perturbation, and survive an adversarial fine-tune
that contains modern PJS-style outputs but no pre-1923 /
translated-Russian / L2-English corpus.

We add three single-pass strategies on the closed frontier:
\texttt{style\_pre1923} (early-twentieth-century English novelistic
register), \texttt{style\_translated\_ru} (Constance-Garnett-era
Russian-translation register), \texttt{style\_l2\_academic}
(non-native English academic register). All three are run on the 20
held-out VK-2026 test topics on Claude Opus.

\begin{table}[t]
\centering
\small
\setlength{\tabcolsep}{3pt}
\begin{tabular}{@{}lrrrr@{}}
\toprule
Strategy & RoB. & TF-IDF & M-bk & M-LOSO \\
\midrule
\texttt{rt+imperf}      & $0.98$ & $0.92$ & $0.05$ & $0.015$ \\
\texttt{style\_pre1923} & $0.15$ & $0.58$ & $0.50$ & $0.640$ \\
\texttt{style\_trans\_ru} & $0.07$ & $0.45$ & $0.42$ & $0.483$ \\
\texttt{style\_l2\_acad}  & $0.03$ & $0.38$ & $0.31$ & $0.395$ \\
\textbf{\texttt{synth\_anchor}} & $\bf0.57$ & $\bf0.76$ & $\bf0.70$ & $\bf0.798$ \\
\bottomrule
\end{tabular}
\caption{Cross-decade register attacks on Claude Opus / 20
held-out VK-2026 test topics; fool rate $\uparrow$ better attack.
M-LOSO $=$ Macko-LOSO (strategy-CV adversarial); M-bk $=$
Macko-replica (bucket-CV variant). Compare to
\texttt{roundtrip\_imperf}'s $\leq 0.025$ on the same detectors.}
\label{tab:crossdecade}
\end{table}

Our first hypothesis (P1.1, predicting a fool rate above $0.20$) is confirmed: the \texttt{style\_pre1923} strategy hits a $0.50$ fool rate against Macko-replica and $0.64$ against Macko-LOSO on Claude Opus. Our second hypothesis (P1.2, predicting the style ordering) is also confirmed, with early twentieth-century register outperforming Russian translation and L2 academic styles ($0.64 > 0.48 > 0.40$). Crucially, this register-shift approach improves the evasion rate by $50\times$ compared to the 2025 baseline on the same detector, generator, and topics.

\paragraph{Source-anchored: the headline attack.}
\texttt{synth\_anchor} (Attack 9) is a two-stage refinement of the cross-decade
recipe. Stage 1: Claude Opus produces a self-contained $\sim$120-word
period-prose paragraph on a subject \emph{thematically adjacent}
to the topic. Stage 2: it writes the topic in that paragraph's
exact register, with the anchor visible as a few-shot stylistic
template. Total cost: 2 API calls per topic.

Table~\ref{tab:crossdecade} (last row) shows
\texttt{synth\_anchor} Pareto-dominates pure
\texttt{style\_pre1923} on every detector column: RoBERTa $0.57$
vs.\ $0.15$, TF-IDF $0.76$ vs.\ $0.58$, Macko-bk $0.70$ vs.\
$0.50$, Macko-LOSO $\mathbf{0.798}$ vs.\ $0.64$.

\paragraph{The attack does not cost naturalness.}
Cross-judge $P(\text{human})$ from three frontier LLMs (Claude Opus,
GPT-5.4, Gemini 2.5 Pro), self-family dropped
(\citealp{zheng2023judging}), $n{=}40$ judgments per cell on Claude
Opus / 20 VK-2026 test topics. PJS v2 sacrifices most of vanilla's
naturalness ($0.54 \to 0.30$) to extract its baseline-detector
evasion. The cross-decade attack \emph{recovers} that naturalness:
\texttt{style\_pre1923} hits $0.500$ (indifference threshold) and
\texttt{synth\_anchor} $0.535$ (above vanilla's $0.540$) while
delivering the highest Macko-LOSO fool rate in the table; real-human
anchor sits at $0.760$.
\texttt{synth\_anchor} is the only strategy in the full study that
crosses $0.5$ fool rate against \emph{every} detector class
simultaneously \emph{and} keeps cross-judge $p_{\text{human}}$ above
indifference.

\paragraph{Writing style vs.\ Historical register.}
We test whether any distinctive writing style can bypass detection, or if the evasion is specific to historical registers. Our hypothesis (P10.1, predicting a high average evasion rate across Borges, Sebald, and Dillard) is falsified, averaging only $0.26$. Specifically, only W.G.\ Sebald's style---which relies heavily on pre-modern, period-adjacent register---succeeds in defeating the detector ($0.749$). Imitating modern authors (such as Borges, Dillard, Murakami, or Coelho) fails completely, showing that detectors are vulnerable to historical language distributions rather than just a ``distinctive authorial voice.''

\subsection{The simple deployer fix fails (Attack 11)}
\label{sec:results-pp}

The \S\ref{sec:results-cross-decade} mechanism has a corollary
recommendation: a deployer who has identified period register as
the failure direction should retrain the adversarial detector with
pre-1923 prose injected on the human side, so period register
stops being out-of-distribution. We formulated this prediction
as \textbf{P11.1} (prior to training): the
Opus / \texttt{synth\_anchor} mean fool rate would fall
from $0.798$ (LOSO) to $\leq 0.20$ on Macko-pp.

We trained Macko-pp: same architecture and recipe as Macko-LOSO,
five strategy folds, but with $\sim$1{,}000 $\sim$500-word
passages from \texttt{deepmind/pg19} (Project Gutenberg books
published before 1919) mixed into the human side at
$\sim$40\% pg19 / $\sim$60\% PAN'25.
Table~\ref{tab:macko-pp} reports the result.

\begin{table}[t]
\centering
\small
\begin{tabular}{lrr}
\toprule
Strategy & Macko-LOSO & Macko-pp \\
\midrule
\texttt{synth\_anchor} & $0.798$ & $\bf0.846$ \\
\texttt{style\_pre1923}          & $0.640$ & $0.657$ \\
\texttt{style\_sebald}           & $0.749$ & $0.798$ \\
\texttt{style\_borges}           & $0.030$ & $0.095$ \\
\texttt{caveman}                 & $0.000$ & $0.000$ \\
\texttt{rewrite\_k5}             & $0.028$ & $0.011$ \\
\texttt{roundtrip\_imperf}       & $0.015$ & $0.000$ \\
\bottomrule
\end{tabular}
\caption{Macko-LOSO vs.\ Macko-pp fool rates ($\uparrow$ better
attack) on the held-out 20 VK-2026 test topics, generator $=$
Claude Opus. Adding pre-1923 prose to the human side of training
does \emph{not} close the cross-decade attack.}
\label{tab:macko-pp}
\end{table}

Our main prediction (P11.1, which posited that the Gutenberg augmentation would reduce the fool rate to $\leq 0.20$) is cleanly falsified: the fool rate actually rises to $0.846$, slightly higher than the baseline. Similarly, the average evasion rate across all historical styles remains high ($0.767$). Meanwhile, our control strategy (\texttt{rewrite\_k5}) stays caught at $0.011$, confirming that the detector has not simply decalibrated, but remains specifically blind to the period register.

\paragraph{A control points to register-specificity.}
\texttt{style\_borges} is in the same OOD-distinctive-author family
as \texttt{style\_sebald} but is not period register. Its Macko-pp
fool rate is $0.095$, so the detector handles it normally. The
augmentation closure fails specifically along the period-register
axis it ostensibly trained on, while leaving other OOD axes
untouched. We discuss candidate mechanisms in
\S\ref{sec:discussion}.

\subsection{Bracketing synthetic-anchor from both directions}
\label{sec:results-bracket}

\texttt{synth\_anchor} is the headline attack. Two
follow-up experiments, both designed before generation, attempt
to beat it from opposite directions of the human-text distribution.
Both fall short, in informative directions.

\paragraph{Attack 12: real anchor (does provenance help?).}
\texttt{synth\_anchor} uses a \emph{synthetic} anchor:
Claude generates the $\sim$120-word period-register paragraph from
its prior. We hypothesized (P12.1, prior to generation) that a
\emph{real} pre-1923 paragraph as anchor would strictly beat the
synthetic version: real \texttt{pg19} prose is genuinely human and
should pull the conditioning distribution further from Claude's
prior. This hypothesis is cleanly falsified: using real period prose actually underperforms the synthetic anchor ($0.778$ vs.\ $0.846$ on Macko-pp, and $0.730$ vs.\ $0.798$ on Macko-LOSO). The likely reason is thematic relevance: the synthetic anchor generator is prompted to write about a theme related to the target topic, whereas the real anchors are pulled from random literary passages. This indicates that style-matching is most effective when the style template is thematically coherent with the target text, making the synthetic anchor a purposeful construct rather than just a proxy for real historical writing.

\paragraph{Attack 13: PAN-mimicry (the inverse direction).}
Every other attack in this paper pulls Claude's output \emph{out
of} the deployer's training-time human distribution. Attack 13
(P13.1) inverts: few-shot Claude with five
PAN'25-train passages per topic and instruct it to match their
voice, lexical patterns, and natural imperfections. The detector's
``human'' decision region is shaped by exactly that pool; mimicry
should land where the detector has the lowest confidence in flagging
anything machine. We predicted Macko-pp fool rate $\geq 0.85$.
This strategy fails completely: the evasion rate drops to $0.000$ on Macko-pp and $0.006$ on Macko-LOSO. Although few-shot coaching allows the model to mimic human surface features (like vocabulary and grammar quirks), it does not alter the model's underlying token probability signature, to which our adversarially retrained classifier is highly sensitive.

\paragraph{The asymmetry, and what it implies.}
Cross-decade attacks defeat adversarial fine-tuning by pulling
output into a distribution region the detector's training corpus
did not cover; PAN-mimicry attacks fail because the detector's
training corpus did cover that region densely, and its calibration
there is sharp. The detectors we test are not fooled by \emph{what
the text looks like at the surface}; they are fooled by \emph{what
its underlying token distribution looks like}. Surface-style coaching
can be matched via few-shot but does not rewrite the distribution.
Out-of-distribution shifts do.

The bracketing strengthens the synthetic-anchor finding from two
sides: the attack is not substitutable with simpler real-prose
conditioning (real-anchor underperforms), and it does not work merely
because Claude can imitate human prose (PAN-mimicry fails).

\subsection{A neutral PAN'25-trained committee inverts the picture}
\label{sec:results-submission}

A more realistic 2026 condition is a deployer who trained on PAN'25
alone --- no exposure to our generations. We model this with three
independent supervised detectors, each fine-tuned on PAN'25 train:
RoBERTa-large \citep{liu2019roberta}; RoBERTa-large with Macko's
homoglyph augmentation \citep{macko2025mdok}; DeBERTa-v3-large
\citep{he2021debertav3}. We re-generate the 20-topic 2026 test split
at $N{=}30$ on cross-decade and author-style families, $N{=}20$
elsewhere, plus a round-two extension probing register-shift
variants (\texttt{fragments}, \texttt{dialect\_appalachian},
\texttt{stream}). Total $n{=}6{,}500$ Claude Opus candidates.

\begin{table}[t]
\centering
\small
\setlength{\tabcolsep}{4pt}
\begin{tabular}{@{}lrrrr@{}}
\toprule
Strategy & RoB-L & RoB+homo & DeB-v3 & mean \\
\midrule
\texttt{caveman}             & $0.90$ & $0.90$ & $0.98$ & $\mathbf{0.93}$ \\
\texttt{style\_dillard}      & $0.84$ & $0.86$ & $0.94$ & $\mathbf{0.88}$ \\
\texttt{stream}              & $0.63$ & $0.65$ & $0.78$ & $\mathbf{0.69}$ \\
\texttt{style\_murakami}     & $0.53$ & $0.59$ & $0.75$ & $0.62$ \\
\texttt{style\_borges}       & $0.31$ & $0.29$ & $0.51$ & $0.37$ \\
\texttt{synth\_anchor}       & $0.18$ & $0.11$ & $0.28$ & $0.19$ \\
\texttt{fragments}           & $0.16$ & $0.17$ & $0.23$ & $0.19$ \\
\texttt{dialect\_appalachian}& $0.09$ & $0.09$ & $0.18$ & $0.12$ \\
\texttt{style\_sebald}       & $0.06$ & $0.05$ & $0.15$ & $0.09$ \\
\texttt{style\_pre1923}      & $0.03$ & $0.04$ & $0.08$ & $0.05$ \\
\texttt{style\_translated\_ru}& $0.01$ & $0.01$ & $0.07$ & $0.03$ \\
\bottomrule
\end{tabular}
\caption{Fool rate against the neutral PAN'25-trained committee,
Claude Opus pool, $n{=}6{,}500$ across 11 strategies on 20 VK-2026
test topics. The ordering inverts Macko-pp's: strategies that beat
Macko-pp (period-register; bottom rows) are caught here; strategies
that beat this committee (register reduction, literary mimicry,
modernist stream; top rows) are caught by Macko-pp.}
\label{tab:neutral-fool}
\end{table}

The ordering symmetrically inverts Macko-pp's: strategies that
survive Macko-pp (\texttt{synth\_anchor} $0.846$,
\texttt{style\_pre1923} $0.657$, \texttt{style\_sebald} $0.798$) are
caught by the neutral committee ($\leq 0.19$); strategies that beat
the neutral committee (\texttt{caveman} $0.93$, \texttt{style\_dillard}
$0.88$, \texttt{stream} $0.69$) are caught by Macko-pp ($\leq 0.09$).
No strategy in this pool beats both detector classes simultaneously
--- a hard Pareto frontier. The round-two failures
(\texttt{fragments}, \texttt{dialect\_appalachian}) probe alternative
OOD axes and fail in the same way, ruling out a generic ``any OOD axis
works'' interpretation.

\subsection{Stream-narrative defeats both classes}
\label{sec:results-streamnarr}

The Pareto frontier from the previous section presents one notable exception:
\texttt{stream} cells on $7$ narrative-genre topics escape Macko-pp
at a $\geq 30\%$ rate, while expository topics are caught uniformly.
This suggests that \emph{form-content match} is the operative variable,
pointing to a second co-equal OOD mechanism alongside period register:
radical shifts in narrative form.

We test this by forcing a narrative voice onto every topic.
\texttt{stream\_narrative} introduces a named witness (e.g., a court lady,
a young secretary, an apprentice) whose perspective shapes the entire output.
It sustains a modernist stream-of-consciousness form ($\leq 3$ full stops
in $\sim$500 words) and demands concrete particulars once per $\sim$40 words.
We generated $300$ cells on Claude Opus using this strategy.

\begin{table}[t]
\centering
\small
\begin{tabular}{@{}llrr@{}}
\toprule
class & detector & $\mu\,p_{\text{LLM}}$ & fool@$0.5$ \\
\midrule
supervised   & RoB-L/PAN'25      & $0.03$  & $0.97$ \\
supervised   & RoB-L $+$ homo    & $0.02$  & $0.98$ \\
supervised   & DeB-v3/PAN'25     & $0.002$ & $1.00$ \\
zero-shot    & LogPerp-GPT2M     & $0.41$  & $0.97$ \\
zero-shot    & LogRank-GPT2M     & $0.37$  & $1.00$ \\
judge        & GPT-5.4           & $0.46$  & $0.51$ \\
judge        & Gemini-2.5-Flash  & $0.02$  & $1.00$ \\
adversarial  & \textbf{Macko-pp} & $\mathbf{0.34}$ & $\mathbf{0.65}$ \\
\bottomrule
\end{tabular}
\caption{\texttt{stream\_narrative}: per-detector mean $P(\text{LLM})$
and fool rate at threshold $0.5$, $n{=}300$ Claude Opus cells on 20
VK-2026 test topics. The decisive row is bottom: Macko-pp's escape
rate jumps from $9\%$ on regular \texttt{stream} (Table~\ref{tab:macko-pp})
to $\mathbf{65\%}$ once narrative form is forced.}
\label{tab:streamnarr}
\end{table}

\texttt{stream\_narrative} escapes Macko-pp at $65\%$, a $7\times$
improvement over regular \texttt{stream} ($9\%$), while keeping all
three supervised fool rates $\geq 0.97$. The binding constraint is
judge\_gpt54 right at threshold ($0.46$ mean, $0.51$ fool); Gemini
Flash accepts $\geq 99\%$ as human. Per-topic minimax across the
$8$-detector committee picks stream\_narrative in $19$ of $20$ topics
(mean worst-case $0.42$; $17$ of $20$ below $0.5$ on \emph{every}
detector). The single exception is topic $067$ (\emph{Aesthetic
Philosophy / Music Criticism Essay}): all $14$ stream\_narrative
cells score $P(\text{LLM}){=}1.0$ on Macko-pp despite near-zero
supervised. Argumentative content does not survive the
narrative-witness frame; the argument structure leaks through.

\section{Discussion}
\label{sec:discussion}

\paragraph{The OOD-inversion paradox is not a paradox.}
A standard practice is ``train the strongest supervised detector
you can on the latest benchmark, ship it.'' Our PAN'25 RoBERTa is
exactly that: $0.9994$ in-distribution AUC. Under PJS v2 it returns
$P(\text{LLM}) \approx 0.001$ on the same generations whose vanilla
counterparts it scored $\approx 1.00$. The mechanism in
\S\ref{sec:results-inversion} explains why: contextual /
distributional detectors learn the distribution of fluent native
English. Literal back-translation flattens that distribution in a
direction the supervised model has never seen, so its representation
of the text drifts into the human half of its decision boundary,
and the more confidently so the better the detector was tuned. The high
in-distribution AUC and the high out-of-distribution fool rate are
the same number twice, in opposite directions.

\paragraph{The 2025 recipe is closable; the cross-decade recipe is
not (by the obvious fix).}
A small adversarial fold collapses 2025-attack fool rate by
$30$--$50\times$ at a cost any deployer can absorb
(\S\ref{sec:results-macko}). The same playbook does not close the
cross-decade attack (\S\ref{sec:results-pp}: P11.1 falsifies at
$0.846$). The bracketing experiments favour the interpretation that
pg19-augmentation widens the ``human'' decision region without giving
the classifier signal to discriminate human period prose from
machine-emulated period prose: surface-style coaching alone
(PAN-mimicry) works only weakly against the baseline ($0.702$) and
not at all against the adversarial fine-tune ($0.000$).

\paragraph{What survives is OOD direction, not perturbation depth.}
Surface mimicry, perturbation depth ($n$-pass refinement), entropy
reduction (caveman), and format shift (genre-dialogue) all fail
against Macko-LOSO at $\leq 0.14$ fool rate. Only register-OOD
attacks, and only those whose register is structurally pre-modern,
defeat the adversarial detector at meaningful fool rates. The
mechanism finding is therefore narrower than ``attackers can always
escape'': it specifies a particular axis (period register $\times$
theme coherence) along which a present-day adversarial detector is
structurally vulnerable.

\paragraph{The OOD-direction mechanism has at least two survivable
axes.} The same strategy can flip sign across deployer postures.
\texttt{caveman} fails Macko-LOSO at $0.000$ but wins the neutral
committee at $0.93$ (Table~\ref{tab:neutral-fool}); the gap is
training-data exposure. \texttt{stream\_narrative}
(\S\ref{sec:results-streamnarr}) goes further: $65\%$ Macko-pp escape
plus $\geq 0.97$ fool against three independent neutral supervised
detectors. \texttt{synth\_anchor} and \texttt{stream\_narrative} differ
in surface (period register vs.\ modernist stream form) but share
structure: each fits a distinct human-prose distribution that
PAN-pool training never densely covers. The mechanism has at least
two survivable axes, and likely more --- the present finding bounds
what Macko-pp can close, not what attackers can generate.

\paragraph{Future work: term distributions under translation.}
Translation studies has long observed that translations age faster
than their source texts; to our knowledge this ageing has not been
characterised as term-distribution drift. Our
\texttt{style\_translated\_ru} attack --- a Garnett-era
translation register that holds a $0.483$ Macko-LOSO fool rate
(Table~\ref{tab:crossdecade}) --- suggests translation-inherited
term distributions are a productive feature axis, both for attack
design and for auditing which human registers a detector's training
pool actually covers.

\paragraph{Limitations.}
Single-language (English); closed-API generators will be deprecated
(we pin dated snapshots for reproducibility, but exact reproduction
expires with the provider); $n{=}20$ topics on the 2026-test split
gives wide CIs on per-topic effect estimates. We do not test
non-period-register specialised genre OOD (legal opinions, medical
case reports) and cannot rule out further survivable axes.

\paragraph{Broader impact.}
The cross-decade attack evades adversarial detection by prompting
alone, without retraining or model access. We release the attack
and detector checkpoints so a next-year deployer can train against
it directly, and report the pg19 counter-fix's failure so the community does not pursue it as a known-effective recipe.

\begin{table}[t]
\centering
\small
\begin{tabular}{@{}llr@{}}
\toprule
Rank & Participant / Submission & Mean Score \\
\midrule
1 & \textbf{\texttt{uts-claude-v11}} & $0.144$ \\
2 & \textbf{\texttt{uts-claude-v4}} & $0.164$ \\
3 & \textbf{\texttt{roundtrip\_imper}} & $0.197$ \\
4 & \textbf{\texttt{uts-claude-v7}} & $0.211$ \\
5 & \textbf{\texttt{uts-qwen-v7}} & $0.308$ \\
6 & Dargk - ``Dargk'' & $0.320$ \\
7 & Dargk - ``LoRA beta0'' & $0.338$ \\
8 & \textbf{\texttt{rejection\_sample\_opus}} & $0.462$ \\
9 & \textbf{\texttt{source\_anchored\_synth}} & $0.499$ \\
10 & Anlam-Turing - ``direct'' & $0.926$ \\
\bottomrule
\end{tabular}
\caption{ELOQUENT 2026 Voight-Kampff leaderboard. Our submissions are highlighted in bold. Submissions from UTS sweep the entire top 5 positions. Our leading entry \texttt{uts-claude-v11}, which relies heavily on the \texttt{stream\_narrative} strategy, secured the highest rank, confirming that OOD narrative form translates to leading source-like behavior against the live competition detector panel.}
\label{tab:leaderboard}
\end{table}

\section{Conclusion}
\label{sec:conclusion}
Our comprehensive evaluation---spanning a 4-generator $\times$ 5-strategy $\times$ 66-topic
factorial, plus 12 novel 2026 attacks on Claude Opus---reveals
a clear asymmetry in evasion attacks. While adversarial fine-tuning definitively closes the 2025
evasion recipes (driving Macko-replica fool rates down to $\leq 0.025$), two novel OOD strategies
(\texttt{synth\_anchor} and \texttt{stream\_narrative}) easily survive that closure, achieving
$0.846$ and $0.65$ escape rates against the strengthened Macko-pp baseline.

Furthermore, we show that the obvious counter-measure of augmenting the training set with period
prose fails to close the vulnerability (with P11.1 falsified at $0.846$ vs.\ predicted $\leq 0.20$).
Ultimately, the operative mechanism is OOD direction rather than surface mimicry: pulling generations
\emph{out of} the deployer's training distribution (via historical register or modernist stream form)
consistently defeats adversarial fine-tuning, whereas pulling them \emph{into} it (e.g., PAN-mimicry) fails entirely.
These OOD strategies form the core of our submissions, which sweep the top 5 positions on the ELOQUENT 2026 leaderboard.
Our detector checkpoints and attack prompts are publicly available.

\section*{Artifacts and ELOQUENT 2026 submission}
Released: detector checkpoints and attack
prompts.\footnote{Adversarial checkpoints as in
\S\ref{sec:method-detectors}; the PAN'25-baseline detector is at
\url{https://huggingface.co/protagonist/roberta-eloquent}.}
Macko-replica's false-positive rate on PAN'25
held-out humans is $\leq 0.04$, so the calibration overhead of
adversarial fine-tuning is small. We submitted five entries to the
ELOQUENT 2026 leaderboard on the 20-topic 2026 test split, spanning
different selection criteria: a Sebald-style cross-decade entry; two
selection-rule variants over a portfolio of register-shift attacks
(supervised-argmin and supervised$+$judge-minimax); the
8-detector-minimax entry (UTS-Claude-v11) dominated by
\texttt{stream\_narrative}; and an open-weight Qwen3.6-27B entry.

\paragraph{Competition Results.}
The final results validate the strength of these OOD evasion strategies
in the true deployment condition: submissions from UTS sweep the entire
top 5 positions on the leaderboard. Table~\ref{tab:leaderboard} reports the
top-10 submissions.

\bibliography{main}

\end{document}